\documentstyle[pra,aps,twocolumn,graphics,epsfig]{revtex}

\begin{document}
\draft
\title{The steady state quantum statistics of a non-Markovian atom laser}
\author{J.J. Hope$^{1,2,\dag}$, 
G.M. Moy$^{2}$, M.J. Collett$^{1}$ and C.M. Savage$^{2}$}
\address{$^{1}$Department of Physics,
University of Auckland, New Zealand \\
$^{2}$Department of Physics and Theoretical Physics,
Australian National University,
ACT 0200, Australia.\\
$^{\dag}$ email: jjh@phy.auckland.ac.nz }
\date{\today}
\maketitle

\begin{abstract} 

We present a fully quantum mechanical treatment of a single-mode 
atomic cavity with a pumping mechanism and an output coupling to a 
continuum of external modes.  This system is a schematic description 
of an atom laser.  In the dilute limit where atom-atom interactions 
are negligible, we have been able to solve this model without making 
the Born and Markov approximations.  When coupling into free space, it 
is shown that for reasonable parameters there is a bound state which 
does not disperse, which means that there is no steady state.  This 
bound state does not exist when gravity is included, and in that case 
the system reaches a steady state.  We develop equations of motion for 
the two-time correlation in the presence of pumping and gravity in the 
output modes.  We then calculate the steady-state output energy flux 
from the laser.
\end{abstract}

\narrowtext

\section{Introduction} \label{sec:Intro}

Since the experimental realisation of a Bose Einstein condensate (BEC) 
in a weakly interacting gas 
\cite{Anderson95,Bradley95,Davis95,Mewes96}, there has been a lot of 
interest in using similar techniques to produce a superior source for 
atom optics experiments, which are limited by the linewidth of thermal 
atomic sources.  It was seen that to produce an atom laser, it was 
simply necessary to generate a BEC and then coherently couple it to 
the outside world 
\cite{Olshanii95,Holland95,Wiseman95a,Spreeuw95,Wiseman95b,Moore96,Guzman96,Janicke96,Borde95,Moy97}. 
 This was first achieved with sodium atoms at MIT by coupling a BEC 
from a magnetically trapped state to an untrapped state using an rf 
pulse \cite{MITExpts}, and has since been repeated with long rf pulses 
\cite{Bloch99} and Raman transitions \cite{Hagley99}.  Although the 
resultant matter waves have been the most monoenergetic source of 
atoms that have yet been produced, a gain-narrowed atom laser produced 
with a continuous pumping mechanism will have a spectral density which 
is orders of magnitude larger.  In this paper we will describe a fully 
quantum mechanical model for such an atom laser which does not make 
the (invalid) Born or Markov approximations, and also does not make 
the mean field approximation.

While a condensate is held in a trap, it is in an eigenstate of the 
system, and is therefore completely monoenergetic.  If the trap is 
suddenly turned off or the condensate is quickly coupled into free 
space, then the resulting wavepacket will have a spread in energies 
due to the momentum spread of the trapped wavefunction.  Such a 
coupling tends to conserve momentum.  Coupling the atoms out slowly 
will tend to preserve the energy of the intial state, so a 
monoenergetic output will be achieved, but this is at the cost of 
reducing the atomic flux \cite{Bloch99,Hagley99,Hope97a}.  A 
continuously pumped laser can have the best of both worlds.  Due to a 
competition between the pumping and the damping, it can produce an 
increasingly narrow energy spectrum in the output as the pumping and 
the output flux are increased.  It is this feature that we wish to 
discover in our atom laser model.  Several theoretical models of a 
continuously pumped atom laser have been produced 
\cite{Olshanii95,Holland95,Wiseman95a,Spreeuw95,Wiseman95b,Moore96,Guzman96,Janicke96,Borde95,Moy97}, 
but they have all made the Born and Markov approximations for the 
output coupling, which have since been shown to be invalid unless the 
output coupling rate is extremely small \cite{Hope97a,Moy97b,Moy99}.  
In this paper we produce and solve a model for a continuously pumped 
atom laser which does not make these approximations.

The early atom laser models are largely distinguished by their choice 
of cooling method, which was either some form of optical cooling 
\cite{Olshanii95,Wiseman95a,Spreeuw95}, or evaporative cooling 
\cite{Holland95,Wiseman95b}.  Evaporative cooling appears to be less 
appropriate for a continuous process, but it is the only method which 
has experimentally reached the quantum degenerate regime in BEC 
experiments.  In all of these schemes, the model for the damping of 
the cavity was the same as that used in the master equation for the 
optical laser.  The resulting equations were therefore very similar to 
optical laser equations.  This means that they could be solved using 
similar methods, and were shown to produce analogous behaviour.  We 
show here that a correct description of the output coupling leads to 
an irreducibly non-Markovian damping, and can lead to different 
behaviour.

In an attempt to produce a more realistic description of the output 
coupling, we modelled a cavity from which atoms were coupled to free 
space via Raman transitions \cite{Moy97}.  This allowed high 
intensities, spatial control and the possibility of giving the atoms a 
momentum kick from the lasers, which gave the outcoupled beam a 
direction.  This was recently achieved by Hagley {\it et al.} 
\cite{Hagley99}.  We also suggested placing the beam in an atomic 
waveguide such as a hollow optical fibre, which is a possible method 
of achieving good spatial properties, and an effectively one 
dimensional output.  Although the rate equations and the 
practicalities of the Raman coupling scheme appeared favourable, it 
was necessary to produce a more complete theory in order to describe 
its full effects.

A quantum mechanical theory for the output coupling from an atomic 
cavity connected to an external field was then developed 
\cite{Hope97a}.  This theory described the dynamics of a BEC which is 
coupled to the outside world, but is not pumped by some continuous 
process.  It was based on the optical input-output formalism developed 
by Gardiner and Collett \cite{Gardiner85}, but it was complicated by 
the dispersive nature of the atomic field as compared to the linear 
energy spectrum for optical fields in free space.  A general solution 
for the dynamics of the cavity and the external fields was presented 
in terms of Laplace transforms, and an analytical solution was found 
for a simple form of the coupling \cite{Moy97b}.

The quadratic nature of the dispersion relations in free space make 
the dynamics of atom-optical output coupling very similar to those of 
photon emission in materials with a photonic band gap.  Near the edges 
of these bands, the dispersion relations are quadratic rather than 
linear, and this causes the model to behave in a non-Markovian way.  
The same qualitative behaviour was found in treatments of this 
system as we found in our purely damped case for the atom 
laser \cite{John94,Vats98}.  

A complete model of an atom laser must also include the effect of 
interatomic interactions and a pumping mechanism.  The effects of 
atom-atom interactions are included implicitly in the output coupling 
models which describe the fields using the nonlinear Schr\"{o}dinger 
equation (NLSE), which includes the effect of s-wave scattering 
\cite{Naraschewski97,Steck97,Zhang98a,Kneer98}.  These models based on 
the NLSE do not consider the effect of the coherences between the 
lasing mode and the external modes, and therefore implictly made a 
Born approximation.  In fact, all mean-field models are essentially 
semiclassical models, as they assume that the atoms in the lasing mode 
can be described by a spatial wavefunction with a predetermined set of 
quantum statistics.  This makes it impossible to fully describe the 
behaviour of the pumping, and the dephasing of the laser mode must be 
added {\it a posteriori}.  The full quantum mechanical description 
allows us to calculate the quantum statistics using a microscopic 
model and physical parameters, and therefore allows us to calculate 
the linewidth of the resulting output.

Unfortunately, interatomic interactions are very difficult to include 
in a full quantum mechanical model.  We present a model of an atom 
laser which includes a pumping mechanism but assumes that atom-atom 
interactions are negligible.  By making this approximation we have 
been able to model the output coupling without making the Born-Markov 
approximation, which is invalid in most physicially relevant parameter 
regimes.  This is an accurate description of very dilute systems, and 
we show that it can be made self consistently with a laser operating 
well above threshold, as the threshold can be less than $10^{2}$ atoms 
in the laser mode \cite{Hope99}.  This restriction is made for 
calculational purposes rather than as a deliberate design criterion.  
The inclusion of the atom-atom interactions in full generality would 
require a multimode description of the intracavity field.  As 
discussed above, the mean-field methods which have been so successful 
in simplifying this procedure cannot be used without destroying the 
very information that we are trying to find.

It may even be possible to use the single-mode approximation for the 
cavity after the system has reached a steady state, if this mode is 
derived self consistently.  At this time, the number of atoms in the 
cavity may be reasonably well defined, and the complicated dynamics of 
the pumping process may be approximated by a linearised master 
equation term, as may be constructed for an optical laser 
\cite{Walls}.

Section \ref{sec:Model} describes our model of a pumped and damped 
single-mode atomic cavity, and descibes the method by which we may 
determine the properties of the output field if we know the dynamics 
of the mode in the cavity.  Section \ref{sec:dampedonly} shows how the 
solution behaves in the absence of pumping.  Section \ref{sec:pump} 
derives the equations of motion for the system in the presence of 
pumping and discusses the physics of the various limits of the 
equations.  We see that we can write the equations of motion for the 
intracavity field in terms of the intracavity operators only, but that 
this leads to non-Markovian equations.  In 
Sec.~\ref{sec:solutionmethod}, we show how these equations can be 
solved to find the two-time correlation of the lasing mode.  Section 
\ref{sec:nograv} shows how the solution in the absence of gravity can 
be found largely analytically, and that it is not self consistent as a 
steady state solution.  Section \ref{sec:gravity} describes the 
features of the energy spectrum of the output of an atom laser in the 
presence of gravity.  In Sec.~\ref{sec:Conc} we discuss the 
possibilities for extending this model.

\section{The model}   
\label{sec:Model}

We model the atom laser by separating it into three parts.  The lasing 
mode is an atomic cavity with large energy spacing, and when it is 
operating in the quantum degenerate regime, it is effectively 
single-mode \cite{Spreeuw95,Moy97}.  We assume that the cavity is 
single-mode, with annihilation (creation) operator $a^{(\dag)}$ and a 
Hamiltonian $H_{s}$.  The external field has a different electronic 
state from the atoms, so the atoms are no longer necessarily affected by 
the trapping potential.  We model the external modes with the field 
operators $\psi^{(\dag)}(x)$ and the Hamiltonian $H_{o}$.  The 
operators $a$ and $\psi(x)$ satisfy the normal boson commutation 
relations.  The coupling between the lasing mode and the output modes 
will be described by the Hamiltonian $H_{i}$.  The pump reservoir is 
coupled to the cavity by an effectively irreversible process.  At this 
stage, we will describe the pump by the Hamiltonian $H_{p}$, which 
also couples the atoms from a pump reservoir into the system mode.

\begin{figure}
\begin{center}
\epsfxsize=\columnwidth
\epsfbox{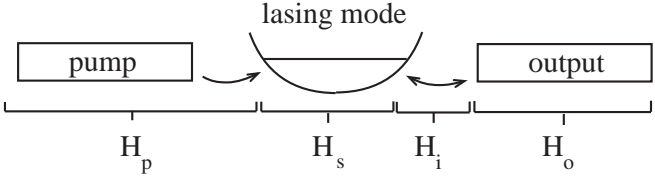}
\end{center}
\caption{Schematic model of the atom laser}
\label{fig:schematic}
\end{figure}

The total Hamiltonian is then written
\begin{equation}
	H_{tot} = H_{p} + H_{s} + H_{i} + H_{o}
	\label{eq:TotalH}
\end{equation}
where
\begin{eqnarray}
	H_{s} & = & \hbar \omega_{o} a^{\dag} a,
	\label{eq:Hsys}  \\
	H_{i} & = & i\hbar \int_{-\infty}^{\infty} 
	    dx \; (\kappa(x,t) \; \psi(x) 
	    a^{\dag}-\kappa(x,t)^{*}\;\psi(x)^{\dag} a)
	\label{eq:Hint}
\end{eqnarray}
and where
\begin{equation}
	\label{eq:pumpCond}
	[H_{o},H_{(s,p)}]=[H_{o},a^{(\dag)}]=[H_{(s,p)},\psi^{(\dag)}(x)]=0,
\end{equation}
as the pump does not directly couple to the external modes.  This is 
described in Fig.~\ref{fig:schematic}.

We then enter the interaction picture, leaving only $H_{i}$.  This 
gives us the interaction Hamiltonian:
\begin{eqnarray}
	\nonumber
	H_{int} &=& i\hbar \int_{-\infty}^{\infty} 
	    dx \; (\kappa(x,t) \; \psi_{I}(x,t) 
	    a_{I}^{\dag}(t)-\\
	&&\;\;\;\;\;\;\;\;\;\;\;\;\;\;\;\;\;\;\;\;\;
	\kappa(x,t)^{*}\;\psi_{I}(x,t)^{\dag} a_{I}(t))	
	\label{eq:hinterfirst}
\end{eqnarray}
where
\begin{eqnarray}
	\nonumber
	 \psi_{I}(x,t)& = &e^{i H_{o} (t-t_{0})/\hbar}\;\psi(x,t_{0}) \; 
e^{-i H_{o} (t-t_{0})/\hbar}
	\label{eq:psiI}  \\
	\nonumber
	a_{I}(t) & = & e^{i/\hbar (H_{s}+H_{p}) 
(t-t_{0})} \;a(t_{0})\; e^{-i/\hbar (H_{s}+H_{p}) (t-t_{0})}
	\label{eq:aI}
\end{eqnarray}
are the system and output field operators in the interaction picture.

We therefore obtain
\begin{equation}
	H_{int} = -i\hbar \left(\xi(t) \;a_{I}^{\dag}(t)-\xi^{\dag}(t) 
	\;a_{I}(t)\right)	
	\label{eq:hint}
\end{equation}
where $\xi(t) = \int dx\; \kappa(x,t) \psi_{I}(x,t)$.

Appendix \ref{app:ioderiv} shows how we can go to the Heisenberg 
picture, and find the result
\begin{equation}
	\psi_{H}(x,t) = \psi_{I}(x,t) - \int_{t_{0}}^{t} ds\;F(x,t-s) 
	a_{H}(s)
	\label{eq:psiH}
\end{equation}
where $\psi_{H}(x,t)$ and $a_{H}(t)$ are the Heisenberg operators 
corresponding to $\psi(x)$ and $a$ respectively, and where
\begin{eqnarray}
	F(x,t,s) & = & [\psi_{I}(x,t),\xi^{\dag}(s)]
	\nonumber  \\
	 & = & \int dy \;\kappa^{*}(y,s) [\psi_{I}(x,t),\psi_{I}^{\dag}(y,s)]
	\nonumber  \\
	 & = & \int dy\; \kappa^{*}(y,s) G(x,t,y,s)
	\label{eq:Fdef}
\end{eqnarray}
where $G(x,t,y,s)$ is the Greens function propagator due to the output 
Hamiltonian, $H_{o}$, only.  These functions can be written in closed 
form for several useful cases including: free space, free space with 
gravity, and a repulsive Gaussian potential.  We may use 
Eq.~\ref{eq:psiH} to calculate any observable in the output field that 
we desire, providing we know the complete history of the system, 
$a_{H}(s)$.  This is a simplified version of the input-output 
relations for an atomic cavity.

\subsection{Output field energy spectrum from $\langle a^{\dag}(t)a(t')\rangle$}
\label{subsec:extractspectra}

When the system is in a steady state, we are not interested in the
external spectrum directly, as it is always growing.  The quantity of
interest is the output energy flux.  We transform our interaction 
Hamiltonian into the basis of the energy eigenstates of the output 
modes:
\begin{eqnarray}
	H_{o} & = & \int dp \;\hbar\:\omega_{p} c_{p}^{\dag} c_{p}
	\label{eq:hodiag}  \\
	H_{i} & = & i \hbar\;\int dp \;\left(\bar{\kappa}(p,t) 
	\:c_{p}\:a^{\dag} - \bar{\kappa}(p,t)^{*}\:c_{p}^{\dag}\:a \right)
	\label{eq:hidiag}
\end{eqnarray}
where $c_{p}$ is the annihilation operator associated with the 
eigenstate of $H_{o}$ that has a position space wavefunction 
$u_{p}(x)$ and energy $\hbar \omega_{p}$, and where $\bar{\kappa}(p,t) 
= \int dx\;u_{p}(x) \:\kappa(x,t)$.

We can then write the output energy flux in terms of the two-time 
correlation of the system.
\begin{equation}
	\frac{d\langle c_{p}^{\dag} c_{p}\rangle}{dt} = 2\; |\bar{\kappa}(p)|^{2}\;
	\Re{\left(\int_{0}^{t} du\;e^{-i \omega_{p} (t-u)} \langle a^{\dag}(t) 
	a(u)\rangle\right)}.
	\label{eq:OutputSpectrum}
\end{equation}
This assumes that at time $t=0$, the output field was in the vacuum 
state.

When the output field is in free space and the only term in $H_{o}$ 
is the kinetic energy, then the eigenstates are the momentum 
eigenstates.  In this case, $\bar{\kappa}(p,t)$ is just the Fourier 
transform of $\kappa(x,t)$.  When there is a gravitational field, 
the eigenstates are Airy functions with a displacement which depends 
on the energy:
\begin{equation}
	u_{p}(x) = {\cal N} \mbox{Ai}[\beta (x - \frac{\hbar\:\omega_{p}}{m\:g})]
	\label{eq:graveig}
\end{equation}
where $\mathcal{N}$ is a normalisation constant, and the length scale 
is given by $\beta = (2 m^{2} g/\hbar^{2})^{1/3}$.  In this case 
$\bar{\kappa}(p,t)$ must be calculated numerically.

We will now derive equations of motion for the system in the absence 
of pumping.

\section{Solution in the absence of pumping} 
\label{sec:dampedonly}

When the cavity is not connected to the pump reservoir, we can 
find the Heisenberg equations of motion for the system operator, and 
then use Eq.(\ref{eq:psiH}) to replace $\psi_{H}(x,t)$ with system 
operators:
\begin{eqnarray}
	\nonumber
	\dot{a}(t) &=& -i \omega_{o} a(t) + \int dx \; \kappa(x,t) 
	\times \\
	&&\;\;\;\;\;\;\;\left(\psi_{I}(x,t) - \int_{t_{0}}^{t} 
	ds\;F(x,t,s) a(s)\right) 
	\nonumber \\
	&=& - i \omega_{o} a(t) - \int_{t_{0}}^{t} ds\;f(t,s) a(s)) 
	+\xi(t)
	\label{eq:adotnopump}
\end{eqnarray}
where
\begin{equation}
	f(t,s) = \int dx \; \kappa(x,t) F(x,t,s).
	\label{eq:fdef}
\end{equation}

For most physical situations, the memory functions $F(x,t,s)$ and 
$f(t,s)$ are simply functions of $(t-s)$, which means that 
Eq.(\ref{eq:adotnopump}) is a Volterra equation of convolution type 
and can be solved using Laplace transforms if the Laplace transform of 
$f$ exists.  This has been solved analytically for simple forms of the 
coupling in earlier work \cite{Hope97a,Moy97b}.  In the limit where 
the decay of the system operator $a(t)$ becomes slow compared to the 
decay of the memory function, the system operator can be taken out of 
the integral.  This is a Markov approximation, and leads to 
exponential decay of the trapped mode.  The general solution 
approaches the Markov solution as the strength of the coupling is 
turned down.  Coupling the atoms out at a finite rate causes a 
significantly non-exponential decay.  The difference between the exact 
and the Markov solutions is examined in detail in parallel work 
\cite{Moy99}.

The most surprising aspect of the exact solution is that it does not 
necessarily go to zero in the long time limit.  If we continuously 
couple a single-mode in a trap to free space, then some of the initial 
population will disperse, but a certain fraction of the initial 
trapped atoms will remain in a non-dispersing eigenstate of the system 
and the coupling.  Even if a momentum kick is given to the atoms as 
they are coupled out, the bound eigenstate exists, although in this 
case it will have a lower occupation.  

The mechanism for the production of this bound eigenstate can be best 
seen in a dressed state picture.  In this picture, the mixed state 
which was originally in free space experiences an attractive 
quasipotential.  Let us find the eigenstate of this system explicitly:

As all of the evolution is coherent and we are not including 
interatomic interactions, the atoms are acting independently in first 
quantised picture, and we may use single particle quantum mechanics to 
describe the eigenstate.  The state of the system is then described 
by a wavefunction
\begin{equation}
	|\Psi\rangle = c_{a} |\psi_{a}\rangle + \int dx \psi(x) |x\rangle
	\label{eq:wavefn}
\end{equation}
where $|\psi_{a}\rangle$ is the lasing mode and $\psi(x)$ is the 
wavefunction outside the trap in the position basis, $\{|x\rangle\}$.

The full Hamiltonian is
\begin{eqnarray}
	H & = & \hbar \omega_{o} |\psi_{a}\rangle \langle \psi_{a}|+ \frac{p^{2}}{2 
	M}
	\nonumber  \\
	 &  & + \int dx\;\left(g(x)|x\rangle \langle 
	 \psi_{a}|+g^{*}(x)|\psi_{a}\rangle \langle x| \right)
	\label{eq:Hsinglep}
\end{eqnarray}
where $M$ is the mass of the atom and $g(x)$ is the shape of the 
coupling.  This Hamiltonian can describe a momentum kick on the atoms 
as they leave the trap by having a Fourier transform of $g(x)$ which 
is not centred around zero.  The trapped eigenstate will be of the 
form
\begin{equation}
	|\psi_{E}\rangle = c_{E} |\psi_{a}\rangle + \int 
	dx\;\psi_{E}(x) |x\rangle.
	\label{eq:eigenstateform}
\end{equation}

The eigenvalue equation $H |\psi_{E}\rangle = E |\psi_{E}\rangle$ 
then leads to the following equations:
\begin{mathletters}
	\begin{eqnarray}
		\int dy \; g^{*}(y) \;\psi_{E}(y) + c_{E} \;\hbar \omega_{o} & = & 
		E \;c_{E},
		\label{eq:eig1}  \\
		c_{E} \;g(x) - \frac{\hbar^{2}}{2M} 
		\frac{\partial^{2}\psi_{E}}{\partial x^{2}}& = & E \;\psi_{E}(x).
		\label{eq:eig2}
	\end{eqnarray}
\end{mathletters}

These are most easily solved in Fourier space, so we take the Fourier 
transform of both sides and define the
Fourier transforms: $\tilde{g}(k)=1/\sqrt{2\pi} \int dx \;g(x) 
\exp (-ikx)$, $\tilde{\psi}_{E}(k)=1/\sqrt{2\pi} \int dx \;\psi(x) 
\exp (-ikx)$.  We then show that the eigenvalue solves the equation
\begin{equation}
	E = \hbar \omega_{o} - \frac{2M}{\hbar^{2}}\;\int dk\; 
	\frac{|\tilde{g}(k)|^{2}}{k^{2} - \frac{2ME}{\hbar^{2}}}
	\label{eq:eigvalue}
\end{equation}
and the corresponding eigenfunction is
\begin{equation}
	\tilde{\psi}_{E}(k) = -\frac{2 M c_{E}}{\hbar^{2}} 
	\frac{\tilde{g}(k)}{k^{2} - \frac{2ME}{\hbar^{2}}}.
	\label{eq:eigstate}
\end{equation}

These equations can be solved analytically for some forms of the 
coupling $\tilde{g}(k)$, and in general they can be solved 
numerically.  For reasonable physical parameters, there is a single 
bound eigenstate with negative energy.  As the coupling is reduced, 
this eigenstate becomes more weakly bound.  If there was a 
gravitational potential in the output field, then we would expect that 
these bound states would become metastable, and would eventually 
decay.  This can be verified numerically \cite{Moy99}.  We would also 
expect atom-atom interactions to destroy this bound state, and in 
current experiments this is certainly the case, as the mean field 
interaction energy is much greater than the initial kinetic energy.

In the following figure, we show that when gravity is added to the 
model, the bound state does indeed decay.  We have chosen a Gaussian coupling 
with width $\sigma_{k}$, which has the form $g(x) = \hbar 
\sqrt{\Gamma} \;(2 \sigma_{k}^{2}/\pi)^{1/4} \;\exp(-(x 
\;\sigma_{k})^2)$.  We chose a weak gravitational field so that the 
short time behaviour is unchanged.

\begin{figure}
\begin{center}
\epsfxsize=\columnwidth
\epsfbox{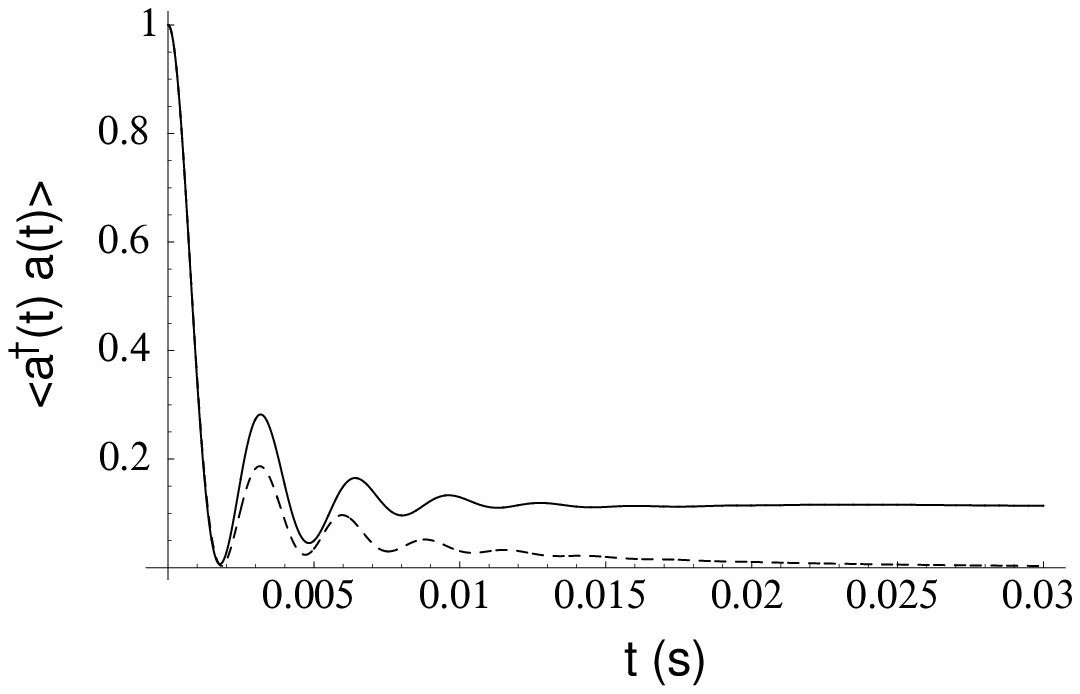}
\end{center}
\caption{This figure shows the decay of the bound state when gravity 
is introduced.  The solid line shows the evolution of the number of 
atoms in the trap without gravity, and it approaches a steady state 
value.  The dashed line shows how it decays in the presence of gravity 
($g=9.8 \sin(\pi/20)$).  The trap atom number $\langle a^{\dag}(t) 
a(t) \rangle $ is normalized so that $\langle a^{\dag}(0) a(0) \rangle 
= 1$.  Parameters are $\Gamma = 10^{6} \mbox{s}^{-2}$, 
$\sigma_{k} = 10^{6} \mbox{m}^{-1}$, $m = 5 \times 10^{-26} 
\mbox{kg}$, $\omega_{o} = 2 \pi \times 123~ \mbox{s}^{-1}$.}
\label{fig:decayonly}
\end{figure}

If we solve Eq.(\ref{eq:eig1}) and Eq.(\ref{eq:eig2}) for this 
example, we find that there is a bound eigenstate of energy $E/\hbar=-4.3 
\times 10^{2}s^{-1}$ with $|c_{E}|^{2}=0.34$.  This means that $0.34$ of 
the original state of the system was in the eigenstate, and the rest 
has gone.  It also means that $0.34$ of that remaining eigenstate will 
be in the trap at any one time.  This means that $0.34^{2} = 0.12$ of 
the initial population of the trap will remain in the trap in the long 
time limit, which is exactly what we see in our simulation.

It is perhaps worth reiterating here that the analysis of this 
non-Markovian system is almost identical to that found in the field of 
photonic band gaps.  Identical population trapping (in this case 
photons rather than atoms) has been calculated using identical 
analytical methods \cite{John94,Vats98}.  There are problems in using 
the Laplace transform method for slightly generalised models, however, 
even though the equations look almost identical.  Particular attention 
must be paid to the abscissa of convergence, as the Laplace transforms 
of many physically interesting kernels do not exist at all.  A 
numerical method based on Laplace transforms will give incorrect 
answers in this case, and alternative methods must be employed.

Gravity is not the only possibility for modelling an output coupling 
which will eventually remove all of the atoms from the trap.  In 
practice, interatomic interactions may also destroy the stabilty of 
this state.  The largest potential seen by the output atoms other than 
gravity will be a repulsion from the trapped atoms.  This is because 
the output atoms will be relatively dilute compared to the atoms which 
remain in the trap.  This can be modelled by an external potential for 
the atoms which will repel them from the trap.  We will assume that 
both of these effects are small so that we can continue to use our 
single-mode approximation for the lasing state.

The existence of a trapped state when there is no pumping will lead to 
concern later, as it will mean that pumping will cause the population 
of the trapped state to increase indefinitely, and no steady state 
will be reached.  

When the model is generalised to include the possibility of pumping, 
the individual Heisenberg operators $a(t)$ in the two-time correlation 
can no longer be calculated, so the same techniques for finding a 
solution do not apply.  In the next section we will derive the 
equation describing the two-time correlation of the cavity field in 
the presence of pumping.

\section{Equations of motion including pumping}  
\label{sec:pump}

If the pump reservoir is sufficiently isolated from the cavity and 
external fields, and the pumping process is designed to be 
irreversible, then we may trace over the reservoir states to produce a 
master equation for a reduced density matrix which describes only the 
cavity and the external fields.  One example of a pumping process 
which can satisfy these requirements can be found in our earlier 
model, where cooled atoms in an excited state are passed over the trap 
containing the lasing mode \cite{Moy97}.  The photon emission of the 
atoms would be stimulated by the presence of the highly occupied 
ground state, and they will make a transition into that state and emit 
a photon \cite{Hope96c}.  For a sufficiently optically thin sample, 
which can be made possible by having a very tight, effectively low 
dimensional trap, the photon is unlikely to be reabsorbed, and the 
process is effectively irreversible.  We choose to model an optical 
cooling process rather than the more experimentally successful 
evaporative cooling process as we are particularly interested in 
designing a continuously pumped system with a steady state.

After we have traced over the pump modes, we will produce a term due
to the effect of the pump in the master equation for the reduced
density matrix $\rho$.  It is important to note that we are not
tracing over the output field modes, so our reduced density matrix
spans the output field as well as the cavity field.  

We derive a pumping term based on an approach similar to that followed 
by Scully and Lamb \cite{Scully67} and found in standard quantum 
optics texts \cite{Walls}.  We model pumping by the injection of a 
Poissonian sequence of excited atoms into the atom laser.  These atoms 
may spontaneously emit a photon and make a transition into the atom 
lasing mode.  Alternatively, they may make a transition into other 
modes of the lasing cavity.  For simplicity, we consider an effective 
two-mode approximation.  To obtain the pumping term, we consider the 
effect of a single atom injected into the atom laser, and then extend 
this to describe the effect of a distribution of atoms.  This gives a 
master equation describing pumping in the number state basis of the 
lasing mode $\rho_{n m} = \langle n|\rho|m\rangle$, where each of 
these elements still include the output field.  This equation can be 
expressed as
\begin{eqnarray}
\left(\dot{\rho}\right)_{\mbox{pump}} &=& r {\cal D} [a^{\dag}] \left( 
n_{s} + {\cal A}[a^{\dag}]\right)^{-1} 
\rho. \label{Eq.MasterEqn3}
\end{eqnarray}
where the superoperators ${\cal D}$ and ${\cal A}$ are defined by
\begin{eqnarray}
{\cal D}[c] &=& {\cal J}[c] - {\cal A}[c], \\
{\cal J}[c] \rho &=& c \rho c^{\dag},\\
{\cal A}[c] \rho &=& \frac{1}{2}(c^{\dag}c \rho + \rho c^{\dag}c).
\label{eq:SuperOps}
\end{eqnarray}

This is the same form as presented for a generic laser master equation 
by Wiseman \cite{Wiseman97a} where $r$ is the rate at which atoms are 
injected into the cavity, and $n_{s}$ the saturation boson number.  In 
our particular model $n_{s}$ depends on the ratio of the probability 
that an atom will spontaneously emit into the lasing mode and the 
probability that the atom will emit into another mode.  This pump 
process is Markovian for the same reasons that damping in optical 
lasers is Markovian, as there are photons being lost from the system.

We may write this pumping term in the number basis, 
$\rho_{n, m}=\langle n|\rho|m\rangle$:
\begin{eqnarray}
\left(\dot{\rho}_{n, m}\right)_{\mbox{pump}} &=& r \left(\frac{\sqrt{n 
m}}{n_{s}+(n+m)/2}\rho_{n-1, m-1} \right. \nonumber \\
\left. \right. &\left. \right. & \left.- \frac{(n+m+2)/2}{n_{s} + (n+m+2)/2} 
\rho_{n, m}\right).
\label{eq:rhoNMdot}
\end{eqnarray}
At this stage, the elements $\rho_{n, m}$ are operators which contain 
the state of the output modes.  For optical lasers, we usually make 
the Born and Markov approximations at this stage.  This is not possible 
for atom lasers except in extreme parameter limits, 
but we will review the results than can be obtained with these 
approximations so that we can compare them with the exact results.

\subsection{Pumping using the Born approximation}

The Born approximation assumes that the density matrix $\rho$ can be 
separated into the product of the reduced density matrix for the 
system $\sigma$, and the reduced density matrix for the output field.  
It further assumes that the output field remains in its original state.  
If we then trace over the output field modes, we obtain a master 
equation for the lasing mode which has the same pumping term as given 
above \cite{Moy99}.  The reduced density now describes only the cavity 
mode, and $\sigma_{n,m} = \langle n|\sigma|m\rangle$ are c-numbers.  
The damping term has the form:
\begin{eqnarray}
\left(\dot{\sigma}\right)_{\mbox{damp}} &=& 
-\int_{0}^{t}du\;[f(u) e^{i \omega_{o} u} 
(a^{\dag}a\sigma(t-u) 
\nonumber  \\
&&\;\;\;- a \sigma(t-u) a^{\dag})+\mbox{h.c.}]
\label{eq:BornDamping}
\end{eqnarray}
where $f(t)$ has been defined in Eq.(\ref{eq:fdef}). 

If we combine the pumping and damping terms, we can generate equations 
of motion for the steady-state atom intracavity number, 
$p_{n}=\sigma_{n, n}$:
\begin{eqnarray}
\label{eq:EOMnumdist}
\dot{p}_{n}(t) &=& \frac{r n}{n+n_{s}} p_{n-1}(t) -
\frac{r (n+1)}{n+n_{s}+1} p_{n}(t)\\
\nonumber
&&- 2n \int_{0}^{t}du\;\Re(f(u) e^{i\omega_{o}t}) p_{n}(t-u)\\
&&+ 2(n+1) \int_{0}^{t}du\;\Re(f(u) e^{i\omega_{o}t}) 
p_{n+1}(t-u).
\nonumber
\end{eqnarray}
where $\Re(g)$ is the real part $g$.

By setting the derivatives to zero and assuming that the functions 
$p_{n}(t)$ approach a constant $p_{n}^{SS}$ as $t$ goes to infinity, 
we obtain a recursion relation 
\begin{eqnarray}
\label{eq:RRnumdist}
0 &=& \frac{r n}{n+n_{s}} p_{n-1}^{SS} \\
\nonumber
&&-\left( \frac{r (n+1)}{n+n_{s}+1} - n \gamma_{BM} \right) p_{n}^{SS}\\
&&+ (n+1) \;\gamma \;p_{n+1}^{SS},
\nonumber
\end{eqnarray}
where
\begin{equation}
	\gamma_{BM} = 2 \int_{0}^{\infty}du\;\Re(f(u) e^{i\omega_{o}u})
	\label{eq:gammadef}
\end{equation}
which gives us the steady state for the atom number distribution 
in the cavity:
\begin{eqnarray}
p_{n}^{SS} &=& {\cal N} \frac{(r/\gamma_{BM})^{n}}{(n+n_{s})!}.
\label{eq:SSnumdist}
\end{eqnarray}
where ${\cal N}$ is a normalisation constant.  This form is almost 
identical to that obtained for the optical laser \cite{Walls}.  The 
distribution looks thermal for $\alpha=r/\gamma < n_{s}$, and for 
$\alpha > n_{s}$, which is called the above threshold regime, it 
approaches a Poissonian distribution with mean atom number $\bar{n}$ 
and variance $V$ given by: 
\begin{eqnarray}
\label{eq:nbar}
\bar{n} &=& \frac{r}{\gamma_{BM}} - n_{s},\\
\nonumber
V &=& \bar{n} +n_{s}.
\end{eqnarray}

Eq.(\ref{eq:psiH}) will give us the properties of the output field if we 
can calculate the two-time correlation of the intracavity field.  We 
proceed by finding the equation of motion for the expectation value of 
the field operator:
\begin{eqnarray}
\left(\frac{d \langle a^{\dag}\rangle}{dt}\right)_{\mbox{pump}} &=& 
\sum_{n=0}^{\infty} \sqrt{n} \dot{\rho}_{n-1, n}
 \nonumber\\
\nonumber
&=& \sum_{n=0}^{\infty} \sqrt{n} \rho_{n-1, n} 
\left(\frac{r}{2n+2n_{s}+1}\right)\\
\label{eq:pumpdadt}
&\approx& \frac{r}{2(\bar{n}+n_{s})+1} \langle a^{\dag}\rangle
\end{eqnarray}
where we have used the fact that the number distribution is localised 
to replace $n$ by $\bar{n}$ in the denominator.  By expanding the 
exact expression as a Taylor series in $1/\bar{n}$, we see that there 
is an error term of the order of $1/\bar{n}^{2}$.  This is awkward, as 
if the first order terms cancel then the linewidth will be of this 
order.  This is in fact what we are hoping to discover, as it will 
mean that the spectrum will become more narrow as we increase the 
pumping of the laser.  If this occurs, then we may approximate the 
magnitude of the off-diagonal elements of $\rho$ as though it were a 
coherent state.  We then find that the expression given in 
Eq.~(\ref{eq:pumpdadt}) is correct to third order.

The analysis which produced this pumping term was only made possible 
by assuming that the cavity atom number becomes localised around some 
mean.  If we do not make the Born approximation, then this fact must 
be taken on faith, and later shown to be a self-consistent solution.  
Furthermore, if we do not make the Born approximation, then we do not 
know at this stage what the mean value $\bar{n}$ will be in terms of 
the other parameters.  Once we accept that the atom number 
distribution peaks around a particular value, however, then we may 
calculate the output spectrum without having to trace over the output 
modes.

\subsection{Linewidth using the Markov approximation}
\label{sec:Markov}

The Markov approximation can be made in certain physical limits 
\cite{Moy99}.  It corresponds to assuming that the decay of the 
reservoir correlation function $f(t)$ is fast compared to the rate of 
change of the density matrix.  We wish to describe the atom laser in 
the parameter regimes where it becomes invalid but we will compare the 
results of this simple calculation to the results of a calculation 
which does not make this approxiamtion.

If we make the Born approximation, trace over the output modes and 
then further assume that a Markov approximation can be made, then we 
can write the damping term of the master equation as: 
\begin{eqnarray}
\left(\dot{\sigma}\right)_{\mbox{damp}} &=& \gamma_{BM} {\cal D}[a] 
\sigma.
 \label{eq:MarkovDamping}
\end{eqnarray}
The derivation of this equation has been examined in detail in other 
work \cite{Moy99}, and it has been assumed to be of this form in 
earlier atom laser models.  It can be generated from 
Eq.(\ref{eq:BornDamping}) by assuming that $\sigma(t-u)=\sigma(t)$, 
and can therefore be taken out of the integral.  The damping 
constant $\gamma_{BM}$ is the same as the one defined in 
Eq.(\ref{eq:gammadef}).

The total equation of motion for the expectation value becomes 
\begin{eqnarray}
\frac{d \langle a^{\dag}\rangle}{dt} &=& 
\left(\frac{r}{2(\bar{n}+n_{s})+1} - \frac{\gamma_{BM}}{2}\right)
\langle a^{\dag}\rangle
 \label{eq:dadtMarkov}\\
\nonumber
&=& -\frac{r}{4 \bar{n}^{2}} \langle a^{\dag}\rangle.
\end{eqnarray}
which has an error term proportional to $r/\bar{n}^{3}$.

The solution to this equation is an exponential decay, and the energy 
spectrum can therefore be shown to be Lorentzian, with a width of 
$\Gamma = r/(4 \bar{n}^{2})$, which is of the order of 
$\gamma_{BM}/\bar{n}$, and will therefore go to zero as we increase 
the pumping of the laser.  This is called gain-narrowing, and it is a 
well-recognised feature of the optical laser, but even in optical 
lasers it is a feature which only exists for certain pumping models.  
If the pump has a significant response time, then an optical laser can 
have a linewidth which scales as $r/\bar{n}$, which is simply 
proportional to the cavity linewidth \cite{Walls}.  Alternatively, a 
well-designed pump can actually lead to subPoissonian statistics 
\cite{Walls}.

Since this model of the atom laser exhibits gain-narrowing when we 
make the Born-Markov approximation, we hope to find that it also 
exhibits gain-narrowing when we solve the model correctly, as it is 
this feature which allows the atom laser to produce output with a high 
spectral density.  In Sec.~\ref{sec:gravity} we shall show that this 
is indeed the case.

\subsection{Correct treatment of pumping}
\label{subsec:pumping}

If we do not make the Born or Markov approximations, but we do assume 
that the trap population is localised around some (at this stage 
unknown) value $\bar{n}$, then we can use Eq.(\ref{eq:pumpdadt}) to 
produce a more general equation of motion for the expectation value 
$\langle a^{\dag}\rangle$:
\begin{eqnarray}
	\frac{\partial}{\partial t}\ \langle a^{\dag} \rangle (t) = 
         (i \omega_{o} + P)\langle 
	    a^{\dag}\rangle (t) + \int dx \; 
	    \kappa^{*}(x,t) \langle \psi^{\dag}\rangle (x,t)
	\label{eq:Dadag}
\end{eqnarray}
where
\begin{equation}
	P=\frac{r}{2(\bar{n}+n_{s})+1}
	\label{eq:Pdef}
\end{equation}
well above threshold.  Remember that we can no longer relate $\bar{n}$ 
directly to the physical parameters of the problem using 
Eq.(\ref{eq:nbar}), which has used the Born approximation.  If the 
solution exhibits gain-narrowing, then the value of $P$ that we use 
will actually determine our value of $n_{s}$.  We are therefore 
producing a solution method which can be solved iteratively to produce 
a self-consistent solution rather than directly generating it.

Under our assumptions the pumping is effectively linear, so we may use 
the quantum regression theorem.  We then recall Eq.(\ref{eq:psiH}), 
and we may immediately derive the following integro-differential 
equation for the two-time correlation:
\begin{eqnarray}
	\nonumber
	\frac{\partial}{\partial \tau} &\langle a^{\dag}(t+\tau) a(t)
	\rangle = (i \omega_{o} + P) \langle a^{\dag}(t+\tau)
	a(t)\rangle \\
	&\:\:\:\;\;\;- \int_{0}^{t+\tau} 
	du \;f^{*}(t+\tau-u) \langle a^{\dag}(u) a(t)\rangle
	\label{eq:TTCEOM}
\end{eqnarray}
where $\tau > 0$, and $f(t)$ has been defined in Eq.(\ref{eq:fdef}).  
We have set $t_{0}$ to zero, and assumed that at this point in time 
there were no atoms in the output field.  In other words, at $t=0$ the 
coupling was switched on, and the expectation values involving the 
normally ordered operators $\psi_{I}(x,t)$ and $\psi^{\dag}_{I}(x,t)$ 
will be zero.

This equation is not sufficient to specify the dynamics of the cavity, 
as it is only a single partial integro-differential equation in a two 
dimensional space.  We also require the integro-differential equation 
for the intracavity number, which we can generate in a similar 
manner.  Well above threshold, we obtain: 
\begin{equation}
	\label{eq:NEOM}
	\frac{\partial}{\partial t}\langle a^{\dag}a\rangle (t) = 
	r \\
	\nonumber
	- \int_{0}^{t} 
	du \;2 Re\{f^{*}(t-u) \langle a^{\dag}(u) a(t)\rangle\}
\end{equation}

These equations are difficult to solve in general, but can be solved 
simply in various limits.  For example, if the kernel $f(t)$ was a 
$\delta$-function then the equations would become local and the 
solution would be a simple exponential.  This is the case for the 
Markovian example shown in Sec.~\ref{sec:Markov} above.  In the 
broadband limit of the optical theory, the function $f(t)$ is 
equivalent to the Fourier transform of a constant, which is exactly a 
$\delta$-function.  Although the broadband limit can be a good 
approximation for the atomic case as well \cite{Moy97b}, the fact that 
atoms will disperse in free space means that the system has an 
irreducible memory, and is non-Markovian.

\subsection{Memory functions}
\label{subsec.memfuns}

Our model for the output involves a change of electronic state via a 
Raman transition, which has recently been achieved 
experimentally\cite{Hagley99}.  From a specific model of this coupling 
we can derive the memory functions $f(t,s)$ and $F(x,t,s)$.  The form 
of $f(t,s)$ in this case has also been derived and discussed in the 
recent work by Jack {\it et al.} \cite{Jack99a}.  In position space, 
the shape of the coupling will be defined by the shape of the lasers 
multiplied by an envelope of the spatial wavefunction of the trapped 
state.  For simplicity of calculation, we will assume that the 
coupling is Gaussian in form, and that there is no net momentum kick 
given to the atoms.  This will allow us to produce analytical forms 
for the memory functions.  Let the coupling be defined by
\begin{equation}
	\kappa(x,t) = \sqrt{\gamma} \left(\frac{2 
	\sigma_{k}^{2}}{\pi}\right)^{1/4} e^{-(\sigma_{k}\;x)^{2}}
	\label{eq:kappadef}
\end{equation}
where $\hbar \sigma_{k}$ is the momentum width of the coupling and 
$\gamma$ is the strength of the coupling.

In the presence of a gravitational field, $V=mgx$, the Green's function 
in Eq.(\ref{eq:Fdef}) can be found as a standard result 
\cite{Schulman81}:
\begin{eqnarray}
	\label{eq:Greensfn}
	G(x,t,y,s)&=&\sqrt{\frac{1}{4\pi i \lambda (t-s)}} \times \\
	\nonumber
	&&\;\;\;e^{({\frac{i(x-y)^{2}}{4 \lambda (t-s)} -\frac{ig(t-s)(x+y)}{4 
	\lambda}- \frac{ig^{2}(t-s)^{3}}{48 \lambda}})}
\end{eqnarray}
where $\lambda = \hbar/(2M)$.

This leads to the following forms for the memory functions:
\begin{eqnarray}
	F(x,t,s) & = & \left(\frac{2}{\pi}\right)^{1/4} \sqrt{\frac{\gamma \:
	\sigma_{k}}{1-4\:i\:\sigma_{k}\:\lambda \:\Delta t}} \times
	\label{eq:Fgrav}  \\
	 &  & \;\;\;e^{-\frac{\Delta t\:g(\Delta t^{2}\:g+6\:x)+i\:(\Delta 
	 t^{4}\:g^{2} + 12\:\Delta t^{2}\:g\:x -12 \:x^{2})\:\lambda 
	 \sigma_{k}^{2}}{12\:\lambda\:(i + 4\:\Delta t\:\lambda \:\sigma_{k}^{2})}}
	\nonumber  \\
	\mbox{and}&&\nonumber\\
	f(t,s) & = & \frac{\gamma \;\;\exp({-\frac{g^{2}\: \Delta 
	t^{2}}{32\:\lambda^{2}\:\sigma_{k}^{2}}}) 
	\;\;\exp({\frac{-i\:g^{2}\:\Delta t^{3}}{48 
	 \:\lambda}})}{\sqrt{1+ 2\:i\:\lambda\:\Delta 
	t\:\sigma_{k}^{2}}}
	\label{eq:fgrav}
\end{eqnarray}
where $\Delta t = t-s$.

If we make $g=0$, then we have the memory functions for coupling into 
free space.  We can see that $f(t)$ will then go as $1/\sqrt{t}$ in 
the long time limit.  The broadband limit may also be found by taking 
$\sigma_{k}\rightarrow \infty$ while $\gamma/\sigma_{k}$ goes to a 
constant.  This limiting case has been examined in detail in previous 
work \cite{Hope97a,Moy97b}.  For a non-pumped cavity with a broadband 
output coupling, the long time limit of the output spectrum has three 
regimes.  In the limit where the coupling is much faster than the free 
field dispersion rate of the atoms, the energy spectrum looks like the 
cavity wavefunction which has appeared in free space.  In the limit of 
coupling that is very slow compared to the dispersion rate, the output 
approaches a Lorentzian.  As the coupling strength is increased from 
one limit to the other, there is a reasonably complex behaviour which 
at first looks like a deformed Lorentzian, and then produces a 
fringe-like structure in the spectrum.  Eventually, the oscillations 
decay and the spectrum begins to look like the cavity wavefunction.

There are no useful approximations which can simplify the full 
calculation.  For the purposes of numerical calculations when $g\neq 
0$, using the broadband limit is actually less tractable than using 
the more realistic memory function.  This is because the integrals 
involving $F(x,t,s)$ and $f(t,s)$ become unbounded in amplitude, and 
their convergence is due to their highly oscillatory nature.  
Retaining the full form of the equations includes the Gaussian 
envelope, which defines a natural upper bound to the integrals.  Since 
the oscillations grow rapid on the same scale as this envelope decays, 
it is probable that the correct solution and the broadband limit are 
qualitatively very similar.

If the two-time correlation is very slowly changing then it may be 
removed from the integral in Eq.~(\ref{eq:TTCEOM}), which is 
equivalent to approximating the kernel to be extremely narrow.  We may 
therefore guess that well above threshold, where we hope to find a 
slowly decaying two-time correlation (or a narrow linewidth), the 
Markov approximation may hold, and the output will be Lorentzian in 
energy.  However, there is also the possibility of other behaviour.  
Even if the amplitude of the two-time correlation decayed slowly and 
exponentially, a fast change in the phase of the solution would not 
only mean that there was a shift in the resonance, but would also 
significantly change the effective decay constant.  There is no 
obvious way to determine whether the coupling will induce such a 
rotation without actually solving the original Volterra equation.  In 
fact, we must solve Eq.~(\ref{eq:TTCEOM}) and Eq.~(\ref{eq:NEOM}) 
self-consistently to be sure that this system will exhibit 
gain-narrowing of the output field.

In the next section, we will show how we may find a solution to 
Eq.~(\ref{eq:TTCEOM}) and Eq.~(\ref{eq:NEOM}), and then in the 
following sections we will calculate the output from a laser with 
$g=0$, and then with a nonzero gravitational field.

\section{Solution methods}
\label{sec:solutionmethod}

\begin{figure}
\begin{center}
\epsfxsize=44mm
\epsfbox{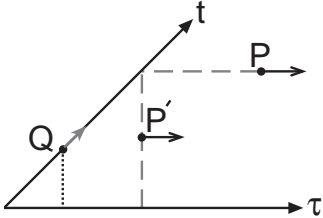}
\end{center}
\caption{The $\tau>0$ half plane of the two-time correlation.  
Equation~(\ref{eq:TTCEOM}) defines the horizontal partial derivatives 
over the entire region, and the domain of the integral corresponding 
to the derivative at point $P$ is shown with a dashed line.  The only 
derivative known at point $P'$ is in the direction shown.  
Equation~(\ref{eq:NEOM}) defines the derivative along the diagonal, 
and the domain of the integral corresponding to the derivative at 
point $Q$ is shown with a dotted line.}
\label{fig:intregion}
\end{figure}

Equation~(\ref{eq:TTCEOM}) and Eq.~(\ref{eq:NEOM}) do not form a 
standard pair of partial integro-differential equations.  The 
derivative in Eq.~(\ref{eq:TTCEOM}) is only defined for $\tau>0$, and 
so we cannot require that the solution obey this equation through the 
whole domain of the integral.  This means that Eqn.~(\ref{eq:TTCEOM}) 
cannot be integrated to find the solution, as we do not actually know 
the derivative at any point.  Since the two-time correlation is 
Hermitian, we can we rewrite the integral so that the domain remains 
in the $\tau>0$ plane, but we still do not have a continuously defined 
derivative along the length of the integral.  This is shown 
graphically in Fig.~(\ref{fig:intregion}).

We proceed by making an ansatz which uses the solution of 
Eq.~(\ref{eq:TTCEOM}) which has been extended into the region $\tau< 
0$.  We then use the $\tau>0$ portion of this solution to substitute 
into the two-time correlation in Eq.~(\ref{eq:NEOM}).  We introduce 
the function $J(t)$, which is the solution of the equation
\begin{equation}
	\frac{\partial J(t)}{\partial t} = (i \omega_{o}+P) J(t) - 
	\int_{0}^{t} dw\;f^{*}(t,w) J(w).
	\label{eq:Jdef}
\end{equation}

This means that
\begin{equation}
	\langle a^{\dag}(t+\tau) a(t) \rangle = \langle a^{\dag} a \rangle 
	(t) \frac{J(t+\tau)}{J(t)}
	\label{eq:S1}
\end{equation}
is a solution of Eq.~(\ref{eq:TTCEOM}) with the correct initial 
condition at $\tau=0$.  We then substitute this result into 
Eq.(\ref{eq:NEOM}):
\begin{equation}
	\frac{d\langle a^{\dag} a \rangle (t)}{dt} = r - \int_{0}^{t} 
	dw\;\langle a^{\dag} a \rangle (w)\;\Re{\left(\frac{2 f(t,w) 
	J(t)}{J(w)}\right)}.
	\label{eq:S2}
\end{equation}

Solving these two equations gives the two-time correlation for the 
lasing mode, from which we may find the properties of the output 
field.  It is only consistent with our linearisation of the pumping if 
the number of atoms in the trap, $\bar{n}$, is stable around the value 
which originally produced the parameter $P$.  Since we require $P$ to 
generate the solution, and $\bar{n}$ is simply the long time limit of 
Eq.~(\ref{eq:S2}), the effective free parameter is $n_{s}$.  This 
threshold parameter must be much smaller than $\bar{n}$, so we search 
for a value of $P$ which gives the result $\bar{n}\gg n_{s}$.

Once it is established that Eq.~(\ref{eq:S2}) is approaching a stable 
steady state, a fast way of finding that steady state is to set the 
derivative to zero, and assume that $\langle a^{\dag} a \rangle 
(w)=\bar{n}$ over the support of the kernel.  This gives
\begin{equation}
	\bar{n} = \frac{r}{\int_{0}^{t} dw\;\Re{\left(\frac{2 f(t,w) 
	J(t)}{J(w)}\right)}}.
	\label{eq:nbarquick}
\end{equation}

There are many numerical methods for solving Volterra equations, and 
since $f(t,s)=f(t-s)$, Eq.~(\ref{eq:TTCEOM}) is of the convolution 
type, which has many specialised methods of solution.  We have found 
that this equation can be solved analytically by a Laplace transform 
method in the absence of gravity and in the broadband limit.  However, 
for the more general forms of coupling that we have examined in this 
paper, even numerical Laplace transform methods become invalid as the 
Laplace transform of the memory function does not always exist.  In 
these cases we require an alternative numerical method.

We transform Eq.~(\ref{eq:Jdef}) into a rotating frame by introducing 
the function
\begin{equation}
	N(t) = J(t)\;\exp(-i \omega_{o} t - P t)
	\label{eq:Ndef}
\end{equation}
and rewriting the equation of motion for $J(t)$ in terms of $N(t)$:
\begin{equation}
	\frac{\partial N(t)}{\partial t} = - 
	\int_{0}^{t} dw\;H(t,w) N(w)
	\label{eq:otherNeom}
\end{equation}
where $H(t,w) = f^{*}(t,w) \exp(-[i \omega_{o} + P](t-w))$.  Since 
$f(t,w)$ is actually just a function of $(t-w)$, this equation is of 
convolution type.  Unfortunately, the numerical methods for finding the 
solution of convolution type Volterra equations depend on 
the simplicity of the kernel for their effectiveness.  We have found 
that the most effective numerical method for solving this equation is 
simply a direct integration of Eq.~(\ref{eq:otherNeom}) using a second 
order algorithm for both the integration and the calculation of the 
integral to find the derivative at each timestep.  Such techniques can 
be found in many common collections of numerical methods 
\cite{NumericalRecipes}.

In the following section, we will examine the case where $g=0$, and 
atoms are simply diffusing away from the trap.

\section{Ideal atom lasers must point down}
\label{sec:nograv}

Real atom lasers will be able to point in any direction and still 
expect a beam of atoms to be emitted.  The three reasons for this are 
repulsive atom-atom interactions which will eject the unconfined 
atoms, the presence of gravity which will accelerate the atoms away 
from the lasing mode, and any momentum kick given to the atoms by the 
coupling process.  We shall show that it is not actually possible to 
produce even an idealised model without either the gravity or the 
atom-atom interactions.

We will do this by attempting to make a model of an ideal atom laser 
without considering complicating factors such as atom-atom 
interactions and gravity.  In this limit, we can use analytical 
methods to solve Eq.~(\ref{eq:TTCEOM}).

If we take the Laplace transform of Eq.~(\ref{eq:TTCEOM}), and then 
use the convolution and derivative theorems, then we obtain
\begin{equation}
	J(t) = J(0)\;{\cal L}^{-1}\left\{\frac{1}{s-(i \omega_{o} + 
	P)+{\cal L}\{f^{*}\}(s)}\right\}(t)
	\label{eq:JLapSoln}
\end{equation}
where ${\cal L}\{f\}(s) = \int_{0}^{\infty} dt \;f(t) \exp(-st)$ is 
the Laplace transform of $f(t)$, and ${\cal L}^{-1}$ is the inverse 
Laplace transform.  

When $g=0$ and we go to the broadband limit, then the memory function 
is
\begin{equation}
	f(t) = |\Omega|^{2} (1-i) \frac{1}{\sqrt{\lambda \:t}} 
	\label{eq:fBB}
\end{equation}
where $\Omega$ is a constant related to the strength of the coupling.  

For this form of $f(t)$, we can find the solution to 
Eq.~(\ref{eq:JLapSoln}) using results from eariler work 
\cite{Moy97b,Moy99}.  We discover that in the long time limit it grows 
exponentially.  The number of atoms in the cavity also grows 
exponentially when this solution is substituted into 
Eq.~(\ref{eq:NEOM}).  This shows that for $g=0$, the steady state 
approximation on which we based our linearisation of the pumping in 
Sec.(\ref{subsec:pumping}) is {\it not self-consistent}.

The exponential growth is not physically realistic, and is merely an 
artifact of the breakdown of our approximation.  In essence, our 
approximation will ignore the depletion of the pump if the 
linearisation is not valid and so in that case we would be modelling 
a pump which can deliver an infinite atomic flux.  The reason for the 
absence of the steady state is clear once we have discovered the 
existence of the trapped state which was described in 
Sec.(\ref{sec:dampedonly}).  Since our trap does not empty when there 
no pumping, it is not unreasonable to expect that a certain fraction 
of the atoms coming into the trap from the pump will enter the trapped 
state.  This will mean that the number of atoms in the cavity would 
continually grow.  We expect that solving this system without 
linearising the pump would lead to a solution that involved a trap 
number which increased linearly.

The absence of a steady state means that this model is not suitable 
for describing an atom laser, and a greater level of detail is 
required in order for the model to behave realistically.  In other 
words, our idealised atom laser model must be more complete than the 
idealised optical laser model.  We also realise that it is the nature 
of the output coupling which must be made more realistic.  A purely 
damped cavity must lose all of its atoms in the long-time limit, or 
there is not really any hope that the pumped atom laser will reach a 
steady state.

Our examination of the trapped state in Sec.(\ref{sec:dampedonly}) 
guides us here, as we know that merely adding a momentum kick to the 
coupling does not remove the trapped state.  It is likely that 
appropriately strong and repulsive interatomic interactions would 
provide the required realism, but we find that that the effects of a 
gravitational field are simpler and easier to model.  In the next 
section, we show that an atom laser in a gravitational field does 
indeed produce a self-consistent steady state.

\section{Output from an atom laser}
\label{sec:gravity}

The output from an atom laser where the output atoms accelerate under 
a uniform gravitational field is a function of many physical 
parameters.  In terms of independent timescales for the dynamics, we 
can control the dispersion rate in free space (by choosing the mass of 
the atoms); the time taken for the output atoms to leave the trap due 
to gravity; the coupling rate between the trap and the output; the 
pumping rate; the threshold pumping rate and the trap frequency.  
Within this phase space, we are constrained by physical limits such as 
the choice of atom and atomic trap.  We are also constrained by our 
operational limits, such as our maximum allowable trap density and our 
desire to be well above threshold.  Finally, we are constrained by 
computational limits, which make it difficult to calculate the 
behaviour of the laser when any of these timescales become radically 
different from the others.  In principle it is possible to discover 
approximations that will work in any of these limits, but each one 
would have to be considered separately.  We will simply examine the 
behaviour of the output for many different pumping and damping rates.  
We will choose the remaining parameters as realistically as possible.

The shape of the coupling $\kappa(x,t)$ is determined largely by the 
spatial wavefunction of the laser mode, and this in turn is a function 
of the trap energy.  We use the trap frequency $\omega_{o}=2 \pi 
\times 123$ Hz, which is near those used in experiments 
\cite{Mewes96}.  We use an atomic mass of $m=5 \times 10^{-26}$kg and a 
gravitational field of $g=9.8 \sin(0.18)$ m s$^{-2}$.  We assume the coupling to 
be Gaussian, with a form given by Eq.~(\ref{eq:kappadef}), and a 
momentum width $\sigma_{k} = 4.4 \times 10^{5}$ m$^{-1}$.

\begin{figure}
\begin{center}
\epsfxsize=\columnwidth
\epsfbox{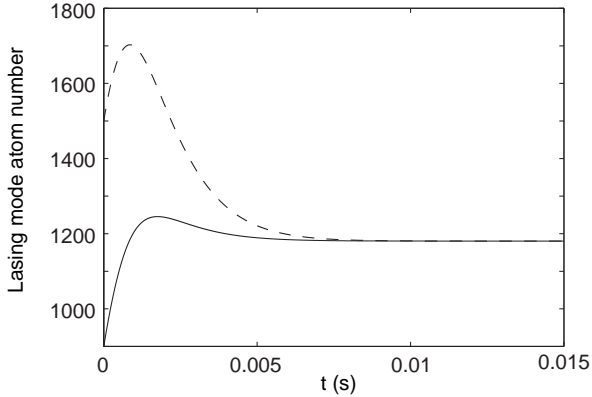}
\end{center}
\caption{This figure shows that the pumped atom laser reaches a steady 
state when the output field experiences a gravitational field.  The 
solid and dashed lines indicate the solution for different initial 
trap numbers.}
\label{fig:Nss}
\end{figure}

Our first calculation will show that the system can reach a 
steady state.  We choose a pumping rate of $r=5.0 \times 
10^{5}$s$^{-1}$, and a threshold cavity number of $n_{s} = 8$.  The 
damping rate is then chosen so that the number of atoms in the trap 
will be much larger than $n_{s}$.  We use $\gamma = 2.0\times 10^{5}$ 
s$^{-2}$, which gives $\bar{n} = 1180 \gg n_{s}$.  We then solve 
Eq.~(\ref{eq:otherNeom}) and use the result to solve 
Eq.~(\ref{eq:NEOM}), which gives us the number of atoms in the trap as 
a function of time.  The results are shown in Fig.~(\ref{fig:Nss}), 
where we can see that after initial transient behaviour, the number of 
atoms reaches a steady state.

The initial transient behaviour is only correct as a first 
approximation, as we are actually determining a self-consistent steady 
state for this system in which the parameter $P$ is stable.  The 
different starting values for $\bar{n}$ are included to demonstrate 
that this solution can be found in a stable fashion.

The two-time correlation derived in this calculation is approaching an 
exponential in the long time limit.  Unlike the Born-Markov solution, 
this exponential has a large rotating term.  This rotation gives the 
output a frequency shift and it also alters the integral in 
Eq.~(\ref{eq:otherNeom}) which in turn affects the rate of decay.  
Even in the strong pumping limit, where the amplitude of the two-time 
correlation is decaying very slowly, the Markov approximation gives an 
incorrect result.

In a counterintuitive way, the fact that the Markov approximation is 
not valid makes it harder to compare the correct calculation to a 
Born-Markov result.  When the memory function is effectively a delta 
function, then the Markov approximation gives the same result in any 
rotating frame, but when the memory function doesn't decay 
instantaneously, then the decay rate and the frequency shift depend on 
the rotating frame in which the Markov approximation is made.  This is 
nothing more than saying that the Markov approximation is not 
self-consistent.  For the purpose of being definite, we will use the 
most ``natural'' version of a Markov approximation, in which we rotate 
at the system frequency $\omega_{o}$.  This gives the linewidth shown 
in Sec.~\ref{sec:Markov} and the damping rate given by the Born 
approximation, Eq.~(\ref{eq:gammadef}).

We now look at what happens when the pumping is increased.  We use a 
damping constant of $\gamma=2.0 \times 10^{4}$ s$^{-2}$, and a 
threshold of $n_{s}=47$, and vary the pumping rate.  As the pumping 
rate increases, the steady state number of atoms in the trap 
increases, and the modulus of the two-time correlation decays more 
slowly.  The energy spectrum of the output flux is proportional to the 
Fourier transform of the two-time correlation through 
Eq.~(\ref{eq:OutputSpectrum}), so we can see that the energy spectrum 
is becoming more narrow as the two-time correlation becomes broader.  
This means that the laser is experiencing gain-narrowing.

In Table~1 we show the results of these calculations.  We give the 
pumping rates $r$ and the resulting mean atom cavity numbers 
$\bar{n}$.  The linewidth of the output energy flux, $\Gamma$, is 
calculated directly from the two-time correlation using 
Eq.~(\ref{eq:OutputSpectrum}).  This is compared to the linewidth given 
by the Born-Markov approximation $\Gamma_{BM}$, which was found from 
Eq.~(\ref{eq:dadtMarkov}).  The spectral shift $\Delta \omega$ is the 
amount by which the correct result is shifted from the Born-Markov 
result.  This spectral shift is seen to be largely a function of the 
damping only, and therefore does not change as we vary the pumping.

We claimed in the introduction that a continuously pumped atom laser 
can have a linewidth much narrower than can be obtained by dropping 
the atoms from the trap after a rapid state change.  In the fast 
coupling limit the output field will simply look like the original 
condensed wavefunction.  More precisely, the output field will have 
the same shape as $\kappa(x)$, which for our example will be a 
Gaussian, given by Eq.~(\ref{eq:kappadef}).  We may proceed to find 
the energy spectrum by changing to the free space energy basis 
($\omega = \lambda k^{2}$).  The resulting energy spectrum is
\begin{equation}
	|\psi(w)|^{2} = \frac{1}{\sqrt{2 \pi \sigma_{k}^{2} \lambda \omega}}\;
	\exp(\frac{-\omega}{2 \lambda \sigma_{k}^{2}}).
	\label{eq:fastlmtNrgSpec}
\end{equation}
which is normalised so that $\int_{0}^{\infty}d\omega \; |\psi(w)|^{2} = 1$.

This spectrum is singular, so a FWHM definition of the width would be 
meaningless.  We shall define the linewidth, $\Gamma_{fast}$, obtained 
by a rapid state change by the equation 
$\int_{0}^{\Gamma_{fast}}d\omega \; |\psi(w)|^{2} = 1/2$.  For the 
parameters used above, this gives us $\Gamma_{fast} = 88$ s$^{-1}$.  
Comparing this linewidth to those calculated in Table~1 shows that our 
atom laser gives an improvement of two to three orders of magnitude in 
spectral density.

\begin{table}[tbp]
	\centering
	\caption{Linewidth as a function of pumping}
\begin{tabular}{|c||c|c|c|c|}
	\hline
	$r$ ($10^{3}$/s) & $\bar{n}$ & $\Gamma$(s$^{-1}$) & 
	$\Gamma_{BM}$(s$^{-1}$) & $\Delta \omega$ (s$^{-1}$) \\
	\hline \hline
	20 & 450 & 2.1 & 0.025 & 0.16 \\
	\hline
	40 & 910 & 1.1 & 0.012 & 0.15  \\
	\hline
	80 & 1800 & 0.56 & 0.0062 & 0.14  \\
	\hline
	800 & $1.8 \times 10^{4}$ & 0.035 & 0.00062 & 0.14 \\
	\hline
\end{tabular}
	\label{tbl:linenarrowing}
\end{table}

\begin{figure}
\begin{center}
\epsfxsize=\columnwidth
\epsfbox{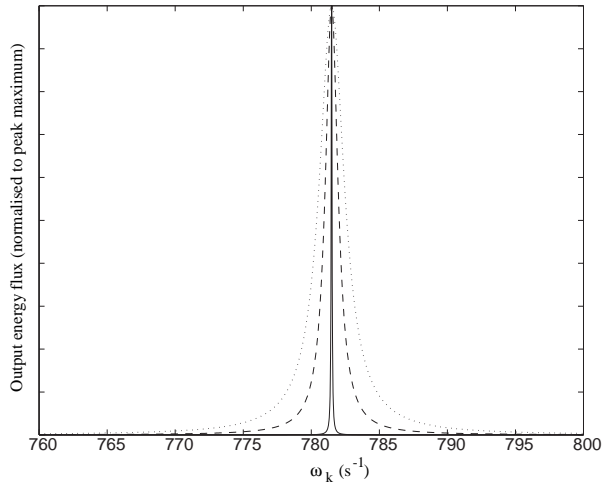}
\end{center}
\caption{The spectra of the output energy flux for three different 
pumping rates.  The solid, dashed and dotted lines represent $r=800$ 
s$^{-1}$, 
$r=80$ s$^{-1}$ and $r=40$ s$^{-1}$ respectively.  Other parameters 
are given in Table~1.}
\label{fig:linenarrowing}
\end{figure}

We plot the spectral flux corresponding to three of these pumping 
rates in Fig.~(\ref{fig:linenarrowing}).  The vertical scale is 
normalised to the peak height for each plot so that the width of the 
spectra can be easily compared.

Although this line narrowing seems to suggest that the linewidth of 
the output can be reduced indefinitely, in practice there will be a 
limit due to technical noise in the pumping, output coupling and the 
laser mode.  A limit due to the finite temperature of the trap has 
been calculated recently by Graham \cite{Graham98}.

\begin{figure}
\begin{center}
\epsfxsize=\columnwidth
\epsfbox{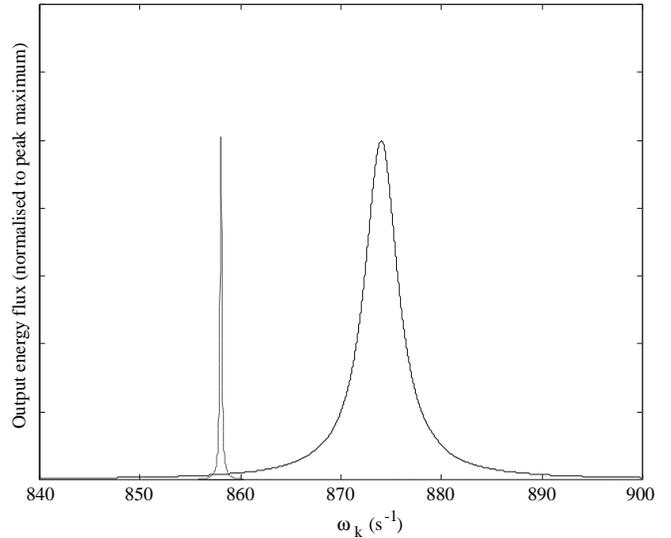}
\end{center}
\caption{The output energy flux spectrum without (solid line) and with
(dashed line) the Born-Markov approximation.  Parameters are $r=5.0 
\times 10^{5} s^{-1}$, $\gamma = 2.0 \times 10^{5} s^{-2}$ and $n_{s} 
= 8$.}
\label{fig:BMcompare}
\end{figure}

From Table~1, we can see that the solution obtained with the 
Born-Markov approximation is significantly different from the correct 
solution.  In Fig.~(\ref{fig:BMcompare}) we plot the output energy 
spectrum with and without the Born-Markov approximation.  We have 
chosen the parameters $r=5.0 \times 10^{5}$ s$^{-1}$, $\gamma = 2.0 
\times 10^{5}$ s$^{-2}$ and $n_{s} = 8$.  This example is well above 
threshold, with a mean atom number of $\bar{n}=1200$.  We scale the 
two solutions so that their peaks are the same height so that they 
can easily be compared.

\section{Conclusions}  
\label{sec:Conc}

We have demonstrated that for some interesting parameter regimes an 
atom laser cannot be modelled by Markovian equations.  Although this 
was shown earlier for a nonpumped cavity \cite{Hope97a,Moy97b,Moy99}, 
there was some hope that the slowly decaying two-time correlation in 
the pumped case would resurrect some form of the Markov approximation.  
We have shown that this can only be true when the coupling is very 
weak, which will tend to make either the atomic flux very weak, or 
else require extremely large atomic densities.

For noninteracting atoms, the non-Markovian equations can be solved 
analytically in the limit of zero pumping, as was the case for the 
pulsed atom laser \cite{MITExpts}.  We have developed a numerical 
method for calculating the output spectrum in the presence of pumping.  
We have found that when the full, non-Markovian dynamics are 
considered, it is necessary to add gravity to the model in order for a 
self-consistent steady state to be produced.

This model does not include atom-atom interactions and therefore only 
works when the atomic field is very dilute, which is consistent with 
our parameters.  The advantage of this model is that we have 
calculated the full quantum statistics of the lasing mode, 
which is impossible with atom laser models which involve nonlinear 
Schr\"{o}dinger equations based on mean-field theory.  Future work 
will involve generalising this atom laser model to include some of the 
effects of interactions without making an initial mean-field 
approximation.  This will allow us to determine the limitations on the 
coherence of a practical atom laser, and thus the limits on their 
interferometric applications.

\acknowledgements

This work was supported by the Australian Research Council, the 
Marsden Fund and the University of Auckland Research Committee.  J.H. 
would like to thank M.Jack, T.Ralph, H.Wiseman and M.Naraschewski for 
their helpful discussions.

\appendix

\section{Deriving the Langevin equation}
\label{app:ioderiv}

In this appendix, we derive Eq.~(\ref{eq:psiH}).  This means that we 
need to find $\psi_{H}(x,t)$, which can be written:
\begin{equation}
	\psi_{H}(x,t) = U_{int}^{\dag}(t,t_{0})\;\psi_{I}(x,t_{0})\; U_{int}(t,t_{0}).
	\label{eq:psiHUint}
\end{equation}
where $U_{int}(t,t_{0})$ is the unitary evolution operator 
corresponding to the interaction Hamiltonian, Eq~(\ref{eq:hint}).  It 
is the identity operator when $t=t_{0}$, and obeys the dynamic equation
\begin{equation}
	\frac{\partial U_{int}(t,t_{0})}{\partial t} = 
	\frac{-i}{\hbar}\;H_{int}(t)\;U_{int}(t,t_{0}).
	\label{eq:uniteom}
\end{equation}

We proceed in a very similar manner to a calculation made by Jack {\it 
et al.} in their work on non-Markovian quantum trajectories 
\cite{Jack99b}.  Let us consider the operator
\begin{equation}
	U(t,t_{0}) = T_{a}\left\{V_{+}V_{0}V_{-}\right\}(t,t_{0})
	\label{eq:udef}
\end{equation}
where $T_{a}$ is a time ordering operator which denotes a time 
ordering on the $a$ operators only, and where we have defined the 
operators
\begin{eqnarray}
	V_{+}(t,t_{0}) & = & \exp\left(-\int_{t_{0}}^{t} ds \;\xi^{\dag}(s) 
	a_{I}(s)\right)
	\label{eq:v+def}  \\
	V_{0} & = & e^{-\int_{t_{0}}^{t} du\;\int_{t_{0}}^{u}dv 
	\;f(u,v) \; a_{I}^{\dag}(u)a_{I}(v)}
	\label{eq:v0def}  \\
	V_{-} & = & \exp\left(\int_{t_{0}}^{t} ds \;\xi(s) 
	a_{I}^{\dag}(s)\right)
	\label{eq:v-def}
\end{eqnarray}
where $f(u,v) = [\xi(u),\xi^{\dag}(v)]$ is the same function defined 
in Eq.~(\ref{eq:fdef}).

Now $U(t_{0},t_{0})$ is clearly the identity operator.  The equation 
of motion for $U(t,t_{0})$ is given by
\begin{eqnarray}
	\frac{\partial U(t,t_{0})}{\partial_{t}} & = & T_{a}\left\{\frac{\partial 
	V_{+}}{\partial t} V_{0}V_{-}\right\}+T_{a}\left\{V_{+}\frac{\partial 
	V_{0}}{\partial t} V_{-}\right\}
	\nonumber \\
	&&\rule{15mm}{0mm}+T_{a}\left\{V_{+}V_{0}\frac{\partial 
	V_{-}}{\partial t} \right\}
	\nonumber  \\
	 & = & -T_{a}\left\{\xi^{\dag}(t)a_{I}(t)\:V_{+}V_{0}V_{-}\right\} 
	 \nonumber \\
	 &&- T_{a}\left\{V_{+}\:\int_{t_{0}}^{t}ds\; f(t,s)\:a_{i}^{\dag}(t) 
	 a_{I}(s) \:V_{0}V_{-}\right\} 
	 \nonumber \\
	 && + T_{a}\left\{-V_{+}V_{0}\:\xi^{\dag}(t)a_{I}(t)\:V_{-}\right\}
	\nonumber  \\
	 & = & \frac{-i}{\hbar} \;H_{int}(t)
	 \;T_{a}\left\{V_{+}V_{0}V_{-}\right\}(t,t_{0})
	\label{eq:Ueom}
\end{eqnarray}
where we used the lemma
\begin{equation}
	T_{a}\left\{[V_{+}(t,t_{0}),\xi(t)]\right\} = 
	\int_{t_{0}}^{t}du\;f(t,u)\;a_{I}(u)\;V_{+}(t,t_{0}).
	\label{eq:comlemma}
\end{equation}

We therefore see that since $U(t,t_{0})$ obeys the same equation of 
motion and has the same initial condition as $U_{int}(t,t_{0})$, then 
it must be the same operator.  We use this alternate form of the 
evolution operator to find $\psi_{H}(x,t)$.

\begin{eqnarray}
	\psi_{H}(x,t) & = & U^{\dag}(t,t_{0})\;\psi_{I}(x,t)\;U(t,t_{0})
	\nonumber  \\
	 & = & \psi_{I}(x,t) + U^{\dag}(t,t_{0})\;T_{a} \left\{ 
	 [\psi_{I}(x,t),V_{+}] \: V_{0}V_{-}\right\} 
	 \nonumber \\
	 & = & \psi_{I}(x,t) + U^{\dag}(t,t_{0}) \; \times
	 \nonumber \\ 
	 &&\;\;\;\;\int_{t_{0}}^{t}ds \; F(x,t,s) \; T_{a} \left\{a_{I}(s) \: 
	 V_{+}V_{0}V_{-}\right\}
	\label{eq:psiHstop}
\end{eqnarray}
where $F(x,t,s) = [\psi_{I}(x,t),\xi^{\dag}(s)]$ is the same function 
as defined in Eq~(\ref{eq:Fdef}).  We then use the lemma
\begin{eqnarray}
	T_{a}\left\{a_{I}(s)\;V_{+}V_{0}V_{-}\right\} & = & T_{a}\{ 
	a_{I}(s) V_{+}(t,s)V_{0}(t,s)V_{-}(t,s) \times
	\nonumber  \\
	&&\;\;\;\;V_{+}(s,t_{0})V_{0}(s,t_{0})V_{-}(s,t_{0}) \}
	\nonumber \\
	&=& U(t,s)\;a_{I}(s)\;U(s,t_{0})
	\label{eq:aislemma}
\end{eqnarray}
to reach the final form of Eq.(\ref{eq:psiHstop}):
\begin{eqnarray}
	\psi_{H}(x,t) & = & \psi_{I}(x,t) - 
	\int_{t_{0}}^{t}ds\;F(x,t,s)\;a_{H}(s).
	\label{eq:psiH2}
\end{eqnarray}

\end{document}